\documentclass[aps,prb,twocolumn,floatfix,groupedaddress,amsmath,amssymb,amsfonts]{revtex4-2}
\usepackage{graphicx}
\usepackage{xcolor}

\begin{document}

%\preprint{}

\title{Real-space Green's function approach for x-ray spectra at finite temperature}


% \author{Tun S. Tan, J. J. Kas, and J.J. Rehr}
\author{Tun S. Tan}
\email[]{tunsheng@uw.edu}
\author{J. J. Kas}
\author{J. J. Rehr}
%\thanks{}
\affiliation{Dept. of Physics, Univ. of Washington, Seattle, WA 98195-1560}


\date{\today}

\begin{abstract}
There has been considerable interest in properties of condensed matter at finite temperature, including non-equilibrium behavior and extreme conditions up to the warm dense matter regime. Such behavior is encountered, e.g., in experimental time-resolved x-ray absorption spectroscopy (XAS) in the presence of
intense laser fields. In an effort to simulate such behavior,  we present an approach for calculations of finite-temperature x-ray absorption spectra in arbitrary materials, using a generalization of the real-space Green's function formalism.
%, including finite-temperature corrections to the exchange-correlation potential.
The method is incorporated as an option in the core-level x-ray spectroscopy code FEFF10. To illustrate the approach, we present calculations for several materials together with comparisons to experiment and with other methods.
\end{abstract}

%\keywords{}

\maketitle

\section{Introduction}
% TODO: Compare with High-temperature electronic structure with the Korringa-Kohn-Rostoker Green's function method (PhysRevE.97.053205). Make real-space dependency explicit.
X-ray absorption spectroscopy (XAS) has become an important tool for  studies of  materials in  fields ranging from materials science and chemistry to geophysics and astrophysics \cite{Savin, Sherborne2015, Lin2017, Cho11, Stamm10, Bolis19}. XAS is used extensively   to probe  both local electronic  and structural properties simultaneously at  synchrotron facilities worldwide.
With the development of x-ray free electron lasers (XFELs), XAS experiments have been extended to treat ultra-short femto- to picosecond time-scales and non-equilibrium conditions, with temperatures $T$ up to many thousands of K. These capabilities are important, e.g., for studies of matter in extreme conditions as well as non-equilibrium and dynamic response due to electron-phonon energy transfer, spin-relaxation and reactions, as well as in shocked matter. The extreme conditions include the warm dense matter (WDM) regime, where $T \sim T_F$ (the Fermi temperature) and the density is within an order of magnitude or so of normal conditions, 

Modern theories of optical and x-ray spectra start  from the many-body Fermi’s golden rule and are usually cast at zero temperature. In order to make the calculations computationally tractable, a single-particle or quasi-particle
approximation for the photoelelectron  is often used \cite{MartinReiningCeperley,prendergastxch}.  The quasi-particle approximation  is the basis of the real-space Green's function (RSGF) approach used in the FEFF codes \cite{Rehr09}, which are applicable to systems throughout the periodic table. Our main aim in this work is  to extend the RSGF  approach for XAS  to finite temperature (FT). Secondly we aim  to explore FT and non-equilibrium behavior in a few systems that can be measured experimentally.
%Corrections due to multi-electron excitations can be added {\it ex post facto} in terms of a convolution with the spectral function \cite{Campbell,kas_phcumulant}, which typically leads to some additional broadening.
%Instead of the conventional ground state exchange-correlation functional,
%FEFF10 uses a dynamic, complex valued self-energy based on the Hedin GW approximation.\cite{Hedin-Lundqvist}
%A common approach to evaluate the cross section is to employ ground-state density
%functional theory (DFT) or the real-space multiple-scattering approach (RSMS).
%Exchange-correlation functionals in many DFT implementations are only available
%for ground-states which requires additional corrections to the functionals
%for excited states in the XAS calculation. On the other hand, the RSMS approach %uses
%complex-valued self-energy in place of the ground-state exchange-correlation.
%In order to treat elevated temperatures a finite-temperature (FT) generalization of the theory is required.
Our FT generalization of the RSGF approach requires a number of extensions, since many electronic ingredients, e.g.,
the chemical potential, exchange-correlation potential, quasi-particle self-energy, and mean-free path are  temperature dependent.  Vibrational effects are also temperature dependent and strongly damp the fine-structure in  XAS at high $T$.
At low temperatures compared to the Fermi temperature $T << T_F$   (which is typically a few $\times 10^4$ K), the exchange-correlation potential and self-energy are weakly temperature dependent and  a zero-temperature approximation is often adequate for electronic structure. Vibrational effects become substantial when $T$ is of order the Debye temperature $T_D$ (which is typically a few $\times 10^2$ K). However, in the WDM regime where $T \sim T_F$, an explicit account of temperature dependence becomes necessary, since the exchange-correlation potential varies from exchange- to correlation-dominated behavior with increasing temperature in that regime \cite{KasBlantonRehr19}.
%In this paper, we present the finite-temperature generalization of the %quasiparticle theory of XAS within real-space Green's function framework.
%We derive
At finite-temperature an efficient approximation for the XAS cross-section can be obtained using the RSGF formalism, which is the real-space analog of the KKR band structure approach \cite{sprkkr}. As at $T=0$, the approach is based on the  one-particle electron Green’s function, which is calculated either by matrix inversion or by a well converged multiple-scattering path expansion. Our current implementation builds upon the theory and algorithms used in the FEFF9 code \cite{Rehr09}, and has been incorporated in a new version FEFF10 \cite{feff10ref}. This theory is illustrated here with a number of examples. While the theory is generally applicable to periodic or aperiodic systems alike throughout the periodic table, we focus on applications at  normal densities, where the theory can be tested against recent experiment.
%Then, we highlight out the relevant changes
%required for a finite-temperature calculation our construction of the %muffin-tin potential also includes the temperature correction to the %exchange-correlation
%potential.

The remainder of the paper is organized as follows: in Sec.\ II.\ we review the
quasiparticle XAS theory and present the RSGF formalism.
Sec.\ III.\ treats finite temperature effects and Sec.\ IV.\ presents a number of   examples.    Finally Sec.\ V.\ contains a brief summary and conclusions.

\section{ Finite temperature XAS}
% TODO: Check formulas.
In this section, we briefly summarize the generalization of quasiparticle theory of XAS to finite temperature.
% While the quasiparticle theory ignores multi-electron excitations such as satellites, these corrections can be roughly approximated by an additional energy dependent broadening factor. Alternatively they can be added {\it ex post facto} by a convolution with the temperature dependent  spectral function
%it has been successful at reproducing many experimental results.
Theories of the XAS cross section typically start from the many-body Fermi golden rule
\begin{eqnarray}
\label{eq:xas_many_body}
\sigma(\omega) = 4\pi^2 \frac{\omega}{c} \sum_{F} |\langle I|\hat{D} |F\rangle|^2 \delta(\omega+E_I - E_F),
\end{eqnarray}
where $|I\rangle$ is the many-body ground state of the system with total energy $E_I$, $|F\rangle$ ranges over the many-body final states with energy $E_F$, and $\hat{D}$ is the many-body transition
operator due to the x-ray field,
%Here  $D$
which here is taken to be a
dipole-interaction ${D}=\Sigma_{i,f} d_{ij}\hat{c}^\dagger_f \hat{c}_i$ where $d_{ij}$ are the dipole matrix elements, and $c_i$ and $c_i^\dagger$ are electron annihilation and creation operators respectively.  Here and throughout this paper we  use atomic units $e=\hbar=m =1$, and units of temperature where $k_B=1$, unless noted otherwise.
In order to reduce the calculation to a quasi-particle framework, we make a sudden approximation as in the  $\Delta$SCF approximation, with the final-state rule \cite{Rehr90}.  Since
%Within the independent particle picture, \\
a core electron leaves behind a hole after being excited into the photo-electron state the final photo-electron state $|\phi_f\rangle$ must take into account the interaction with the core hole potential, while a given core-level $|\phi_i\rangle$ is calculated with the initial state configuration. Next the many-particle initial and final states are factored such that $|I\rangle = |\phi_i\rangle   |\Phi^{N-1}_0\rangle$ and $|F\rangle = |\phi_f\rangle  |\tilde\Phi^{N-1}_n\rangle$, where $|\tilde\Phi^{N-1}_n\rangle$ is the $n^{th}$ excited state of the $N-1$ electron system with a core-hole in level $i$, and $|\phi_f\rangle$ is the   photo-electron state calculated in the presence of the core-hole. If one ignores the energy dependence of excitations of the $N-1$ electron system, Eq.\ (\ref{eq:xas_many_body}) yields an effective single- or quasi-particle cross section $\sigma_{1}(\omega)$ contribution from a given core level $i$ given by
\begin{eqnarray}
\label{eqn:xas}
\sigma_{1}(\omega) = 4\pi^2 \frac{\omega}{c} \sum_{f} |\langle i|\hat{d} |f\rangle|^2 \delta_{\Gamma}(\omega+\epsilon_i - \epsilon_f),
\end{eqnarray}
where $\epsilon_i$  and $\epsilon_f$ are the eigenenergies of the quasiparticle initial $|i\rangle$ (deep-core) and final $|f\rangle$ (photo-excited) states, and $\delta_{\Gamma}(\omega)$ denotes a Lorentzian of width $\Gamma$, where $\Gamma$ includes the lifetime and final-state broadening.
More generally this approximation can be corrected by a convolution of the quasi-particle XAS with the core spectral function, defined as $A_c(\omega)=\Sigma_n S_n^2 \delta(\omega-\omega_n)$ where
$S_n^2= |\langle \Phi^{N-1}_0|\tilde\Phi^{N-1}_n\rangle|^2$ is the many-body amplitude overlap integral between initial and final states. This convolution modifies the cross-section by an additional, energy dependent broadening factor \cite{kas_phcumulant}, of unit weight which damps the fine structure by an constant factor, which historically has been denoted by $S_0^2$ \cite{RehrAlbers2000}.

\subsection{Real-space Green's function }

%and the translational symmetry requirement in Korringa-Kohn-Rostoker (KKR) band structure
Formally the retarded single electron Green's function $G^R(\epsilon)$ in a basis of local site, angular momentum states\cite{RehrAlbers2000} $|L,j\rangle=j_L({\bf r})Y_L(\hat r)$ is given by the spectral sum
%$G^R_{L'j',Lj}(\epsilon)=$ at site $j$
\begin{eqnarray}
%G^R_{l'j',lj}(\epsilon) = \sum_{f L
%G^R(\epsilon) &=&  \sum_{Lj,L'j'} |L'j'\rangle G_{L'j',Lj} \langle Lj |, \\
G_{L'j',Lj}(\epsilon) &=&  \langle L'j'| G^R(\epsilon) |L,j\rangle \nonumber \\
 &=& \sum_f \frac{\langle L'j'|f\rangle\langle f|Lj\rangle}
{\epsilon-\epsilon_f + i\delta},
\end{eqnarray}
%\begin{eqnarray}
%G^R_{l'j',lj}(\epsilon) &=& \sum_{f l l'} |l'j'\rangle \frac{\langle %l'j'|f\rangle\langle f|lj\rangle}{\epsilon-\epsilon_f + i\delta} \langle lj|\\
%-\frac{1}{\pi} \textnormal{Im}\ G^R_{l'j',lj}(\epsilon) &=& \sum_{f l l'} %\langle l'j'|f\rangle \langle f|lj\rangle %\delta(\epsilon-\epsilon_f)\nonumber\\
%& &\times  |l'j'\rangle \langle lj|,
%\end{eqnarray}
where $\varepsilon_f$ is the eigenenergy of the final state SCF quasi-particle Hamiltonian $H'=p^2/2 + v_f + \Sigma(\epsilon;T)$, $v_f$ is the effective one electron Coulomb potential in the presence of a screened core hole, and $\Sigma(\epsilon)$ is the dynamically screened FT electron self-energy. A significant advantage of the RSGF   formalism is that it implicitly builds in the  summation over final states $f$, which is a computational bottleneck in wave-function approaches at high energies \cite{Rehr09}.
In FEFF the RSGF approach implicitly includes Dirac-relativistic effects, and the angular momentum index
%and a basis of site, angular momentum states $|L,j\rangle$, with
 $L$ denotes the relativistic angular quantum numbers $L=(\kappa,m)$\cite{RehrAlbers2000}. In the following  temperature dependence will be assumed implicitly, unless otherwise specified.  By inserting the spectral representation of $G^R(\epsilon)$ into Eq.\ (\ref{eqn:xas}) together with Fermi-Dirac occupation numbers, the final states are implicitly summed in the calculation of $\sigma_{qp}(\omega)$ using the solution to the Dyson equation  \cite{Ankudinov98},  as summarized below. Thus
\begin{equation}
    \label{eq:finite_temp_xas1}
    \sigma_{qp}(\omega) = L_{\Gamma}(\omega)*\sigma^{0}_{1}(\omega),
\end{equation}
where $L_{\Gamma}(\omega)$ is a Lorentzian function that accounts for the core-hole lifetime broadening
%via a convolution,
and $\sigma_{1}^{0}$ is the single- or quasiparticle cross-section with no core-hole broadening, given by
\begin{eqnarray}
\label{eq:finite_temp_xas2}
\sigma_{qp}^0(\omega) &=& -4\pi \frac{\omega}{c} \,\textnormal{Im} \sum_{ill'} \langle i|\hat{d}\ G^R_{l0,l'0}(\omega+\epsilon_i)\ \hat{d}^\dagger |i\rangle \nonumber\\
& & \times f_T(\epsilon_i) \big[1-f_T(\omega+\epsilon_i)\big],
\end{eqnarray}
where site 0 is taken to be the absorbing atom, $f_T=1/[\exp(\epsilon-\mu(T)) +1]$ is the Fermi-Dirac distribution, and $\mu(T)$ is the chemical potential of the system. Much of the electronic temperature dependence is due to that in the occupation numbers  $f_T$.
%Additional temperature dependence appears in Debye-Waller factors due to thermal
%vibrations and in the self-energy.

Multiple-scattering of the photoelectron by its  environment is naturally built into the RSGF formalism\cite{Ankudinov98}. The intra-atomic contributions from the absorbing atom $G_c$ can be separated from the scattering atoms $G_{sc}$ into $G^R = G^R_c + G^R_{sc}$. (For simplicity, here and below we drop the subscripts on matrix elements of $G$.) The scattering contribution $G^R_{sc}$ is determined by the free Green's function $G^R_0$, the total scattering matrix $T$, and partial-wave phase shifts $\delta$ for the photoelectron state of a given angular momentum (suppressed) using the matrix inverse solution to the Dyson equation \cite{Rehr90}.
\begin{equation}
    G^R_{sc} = e^{i\delta}\big(1-G^R_0 T \big)^{-1}G^R_{0}e^{i\delta'}.
\end{equation}
For EXAFS energies (typically above about 50 eV beyond an absorption edge), the matrix inverse can be expanded in a rapidly converging series of scattering paths typically with less than 4 legs. However, for XANES, full multiple scattering (FMS) is usually needed, and can be  carried out by a fast matrix inversion algorithm since the basis of relevant angular-momentum, site scattering states is small due to the short mean free path and $l_{max} < 4$.

\section{Finite-temperature effects}
Several effects come into play in the theory of XAS at finite temperatures. First, there are effects of electronic temperature in the initial many-body electronic configuration  of the system due to the temperature dependence of the electron density and chemical potential. Second, the Fermi function cutoff that defines the x-ray ``edge" of the cross-section as in Eq.~(\ref{eq:finite_temp_xas2}) broadens with increasing $T$. Third there are FT effects on the exchange and correlation, both through the exchange-correlation potential $v_{xc}$ and the quasi-particle self-energy $\Sigma(\epsilon)$.  Finally, there are the FT effects of lattice vibrations which give rise to Debye-Waller factors that  strongly damp the XAS fine-structure. These effects are summarized below:

%\subsection{Finite-$T$ ground-state: SCF}
1)  FT SCF: As for the ground state, in order to construct the (scattering) potential $v$, we implement a generalization of the self-consistent field (SCF) method to calculate the Coulomb potentials, electron densities and a  temperature dependent chemical potential. In the RSGF approach in FEFF, the SCF procedure starts with the overlapped atomic electron densities obtained by solving the relativistic Dirac-Fock equations for each atom in the system. From this initial guess of the electron density, a Green's function is calculated, which provides a new density and chemical potential. This procedure is then repeated until self-consistency ($\rho_i=\rho_{i-1}$) is reached to high accuracy, typically in about 20 iterations or so.
% old subsection contour integration
Assuming a frozen core, the valence electron contribution to density of electrons is given in terms of an  integral over energy of the imaginary part of the
Green's function,
\begin{eqnarray}
\rho(r) &=& -\frac{2}{\pi}\int_{E_B}^{\infty} dE\ \textnormal{Im}\ G(r, r, E) f_T(E),
\end{eqnarray}
where $E_B$ is the core-valence separation energy (typically set to -40 eV in FEFF), and the factor of 2 accounts for spin-degeneracy.
\begin{figure}[t]
    \centering
    \includegraphics[width=0.4\textwidth]{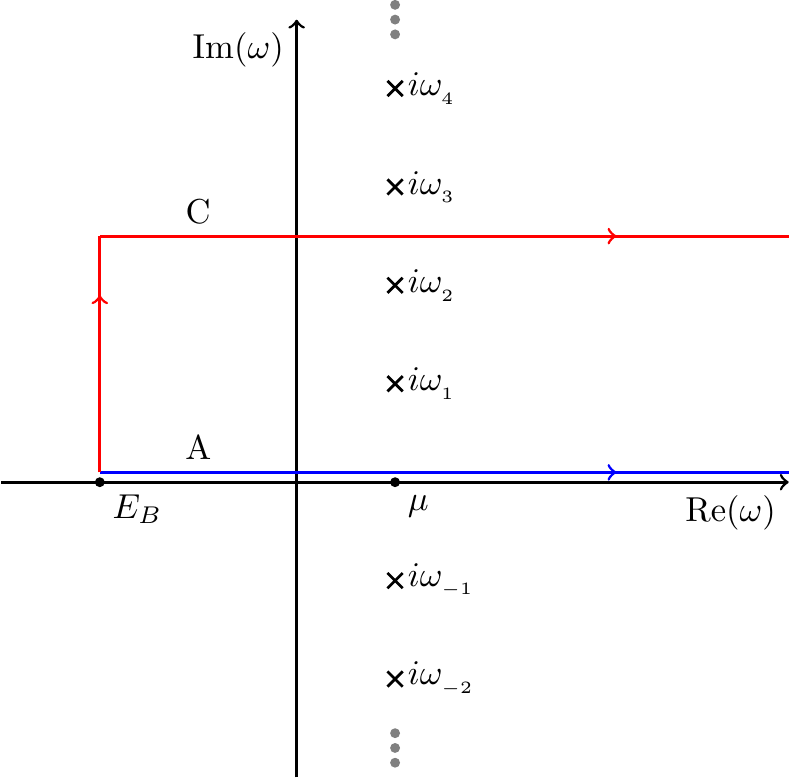}
    \caption{Energy contour for electron density calculation at finite temperature:  $\omega_j = \pi (2j-1)kT$ denotes the imaginary part of the Matsubara poles of the fermi function $f(\omega)$. The integration starts from some core-valence separation   level $E_B$ above the core states but below the valence states. Contour A (blue) represents the original integration path just above the real axis, while contour C (red) represents the integration path in the complex plane.}
    \label{fig:example_contour}
\end{figure}
For computational efficiency at  both zero and finite temperatures, the above integral is carried out in the complex plane where the Green's function becomes smooth. However, there is an added complication due to the possible presence of the Matsubara poles in the Fermi function. In order to include them, we integrate along a  contour C (shown in Fig.~\ref{fig:example_contour}), where the first leg traverses from $E_B$ to $E_B+2\pi i n k T$, then to $\infty+ 2\pi i  n k T$, where $2\pi i n k T$ is halfway between the Matsubara poles. For a given chemical potential, the valence density is then given by
% \textcolor{red}{[I BELIEVE THE SIGN BEFORE THE SUM ON POLES SHOULD BE +. $\int_A - \int_C = \sum_{RES}\rightarrow \int_A = \int_C + \sum_{RES}$]},
\begin{eqnarray}
\label{eq:contourgf}
\rho(r) &=& -\frac{2}{\pi} {\rm Im}\, \Bigg[\int_{C} dE\ \ G(r, r, E) f_T(E),\nonumber \\
&&- 2\pi i kT\sum_{j=1}^{n} G(r,r,z_j)\Bigg],
\label{eqn:rho_exact}
\end{eqnarray}
where $z_j = \mu \pm i\pi (2j-1)kT$ are the Matsubara poles, and $n$ is the number of poles between the contour C and the real axis.
Eq.\ (\ref{eq:contourgf}) also has implicit dependence on the chemical potential through the Fermi function. In turn the chemical potential is found by  enforcing charge neutrality,
\begin{equation}
    N_{el}=\int dr\, \rho(r;\mu(T)),
\end{equation}
where $N_{el}$ is the number of valence electrons in the system, again treating the core-electrons as frozen.
Eqs.~(\ref{eq:finite_temp_xas1}) and (\ref{eq:finite_temp_xas2}) for the cross section contain similar integrals of the Green's function multiplied by the Fermi function. These integrals are dealt with in a similar manner, although the contour is slightly different, in principal extending from $-\infty + i 2\pi  n k_B T$ to $+\infty + i 2\pi  n k_B T$. There is also an additional pole arising from the Lorentzian broadening function $L_{\Gamma}(\omega)$.

2) FT exchange-correlation potentials: Additional temperature dependence comes from that in the exchange-correlation potentials (which affect the initial state densities and potentials). An approach for calculating these exchange-correlation potentials from first principles with a FT cumulant Green's function approach has been discussed by Kas et al.\cite{KasBlantonRehr19} which shows that the potentials cross-over from exchange- to correlation-dominated with increasing $T$. Here, the behavior is  treated by an efficient parameterized extension of DFT fit to Quantum-Monte Carlo calculations (QMC) by Karasiev et al. \cite{KSDT},  which is in good agreement with the results of Kas et al. \cite{KasBlantonRehr19}.

3) FT Self-energy effects:  The temperature dependence of the dynamical quasiparticle self-energy becomes important in calculations of the quasi-particle photoelectron states and hence the transition matrix elements at
energies above an edge. Since full calculations of the FT self-energy\cite{KasRehr17} are computationally demanding, we have used a much more efficient FT COHSEX approximation to the GW self-energy for this purpose \cite{Tan18}.

4) Lattice vibrations: As at $T=0$ Lattice vibrations strongly damp electron scattering at high energy.
%In order to include  these effects, several approximations are needed: 
Approximate calculations  at temperatures up to several times the Debye-temperature can be done by including Debye-Waller factors $\exp(-2\sigma^2 k^2)$ in the Green's function propagator or multiple scattering paths. This can be done  using the correlated Debye model and mean square radial displacements  $\sigma^2(T) \propto T/ T_D$, where $T_D$ is the Debye temperature.
%These methods are reliable up to several times $T_D$. 
However, this approach becomes inapplicable in the near edge regime (where symmetry breaking occurs) and at high temperatures $T>>T_D$, where the quasi-harmonic approximation breaks down. %when states below the zero-temperature
   %edge appear in the FT edge-structure.
   For  those cases a configurational average is called for.
At temperatures well above $T_D$, one can  use finite-$T$ molecular dynamics (MD) to obtain a temperature dependent configurational average of the spectrum. Closer to zero temperature, zero-point motion can also  be included via  Quantum-Monte-Carlo sampling.

%In the next Section we  illustrate our approach for XAS at elevated temperatures  with a number of examples.

\section{Calculations}

\subsection{Simple Metal: Al}
\begin{figure}[h]
        \centering
        \includegraphics[width=0.40\textwidth, height=0.40\textwidth]{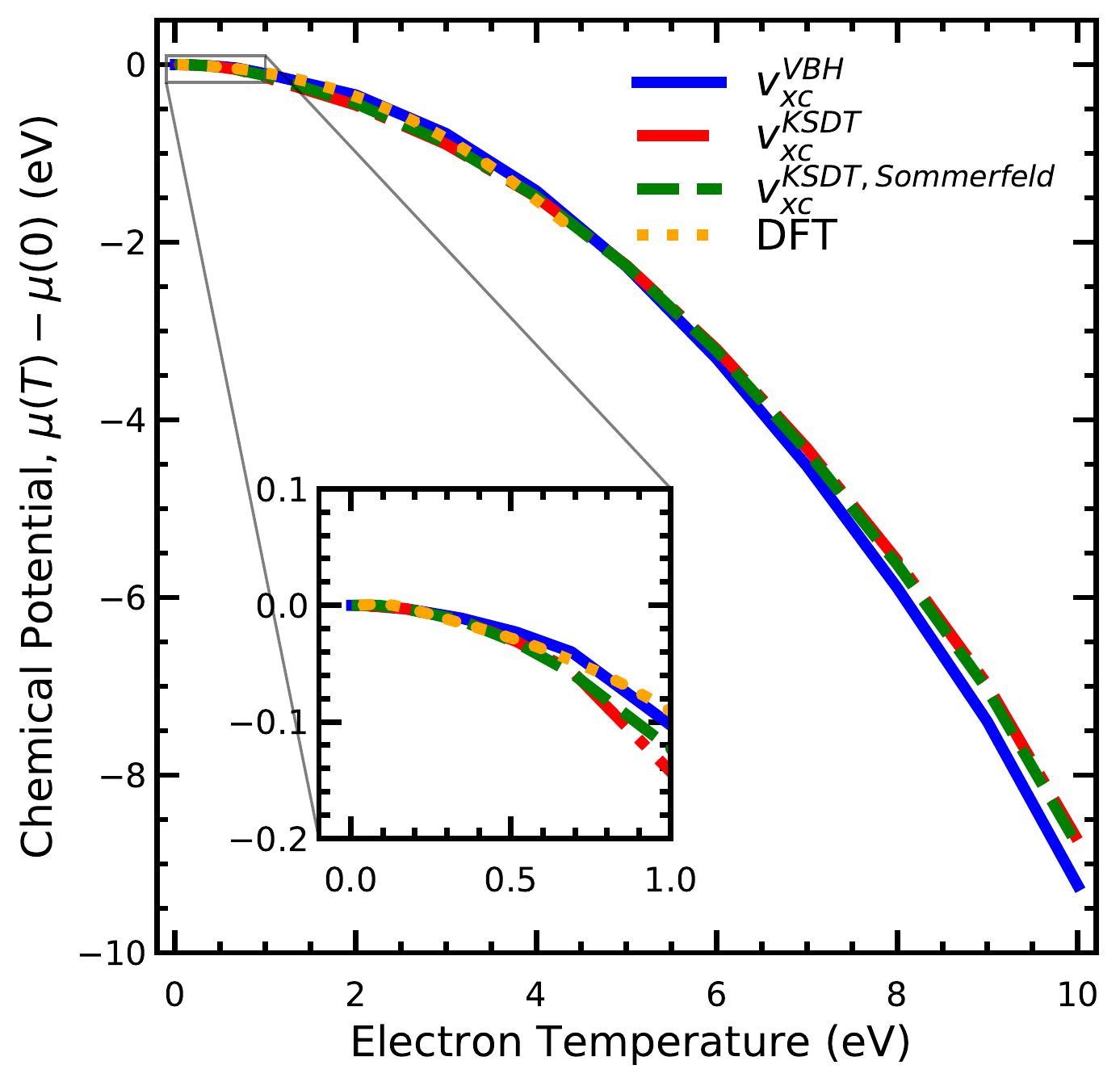}
        \caption{Chemical potential shift for aluminum, $\mu(T)-\mu(0)$ versus electron temperature, $T_e$ using the ground state von Barth-Hedin   \cite{Barth72} (blue) and finite $T$ KSDT \cite{KSDT} (red) exchange-correlation potentials. For comparison the $T=0$ K DFT calculation \cite{Zhibin08} (orange) is also shown up to $T =$ 4 eV. The result with the Sommerfeld expansion (green)  is shown for KSDT exchange-correlation potential. The inset shows the chemical potential shift up to $T = $ 1 eV.
}
        \label{fig:al_mu}
    \end{figure}

    Aluminum (fcc Al) is a prototypical nearly-free-electron system for testing electronic structure calculations. The electronic density of states (DOS) for Al in the conduction band is similar to that for a free electron model, with a square root like dispersion at the bottom of the band. We show  the chemical potential shift vs electron temperature $T_e$ for aluminum up to $T_e$ = 10 eV (1 eV = 11604 K) in Fig.\ \ref{fig:al_mu}.
    For $T=0$ we used the  von Barth Hedin  ($v_{xc}^{VBH}$) exchange-correlation potential \cite{Barth72} and and for $T>0$  the temperature dependent approximation of KSDT ($v_{xc}^{KSDT}$) \cite{KSDT}.
    %We have verified that the temperature correction to the chemical potential is negligible for $T << 1$ eV.
    Our $v_{xc}^{VBH}$ chemical potential agrees well with the calculations of Lin et. al \cite{Zhibin08} up to $T=$ 4 eV, and is reasonably accurate up to about 10 eV, consistent with the $T^2$ behavior of the Sommerfeld expansion. %\textcolor{red}{[SHOW SOMMERFELD]}
Remarkably the Sommerfeld-expansion remains a good approximation even at very high $T$ of order several eV.
    \begin{figure}[b]
        \centering
        \includegraphics[width=0.4\textwidth]{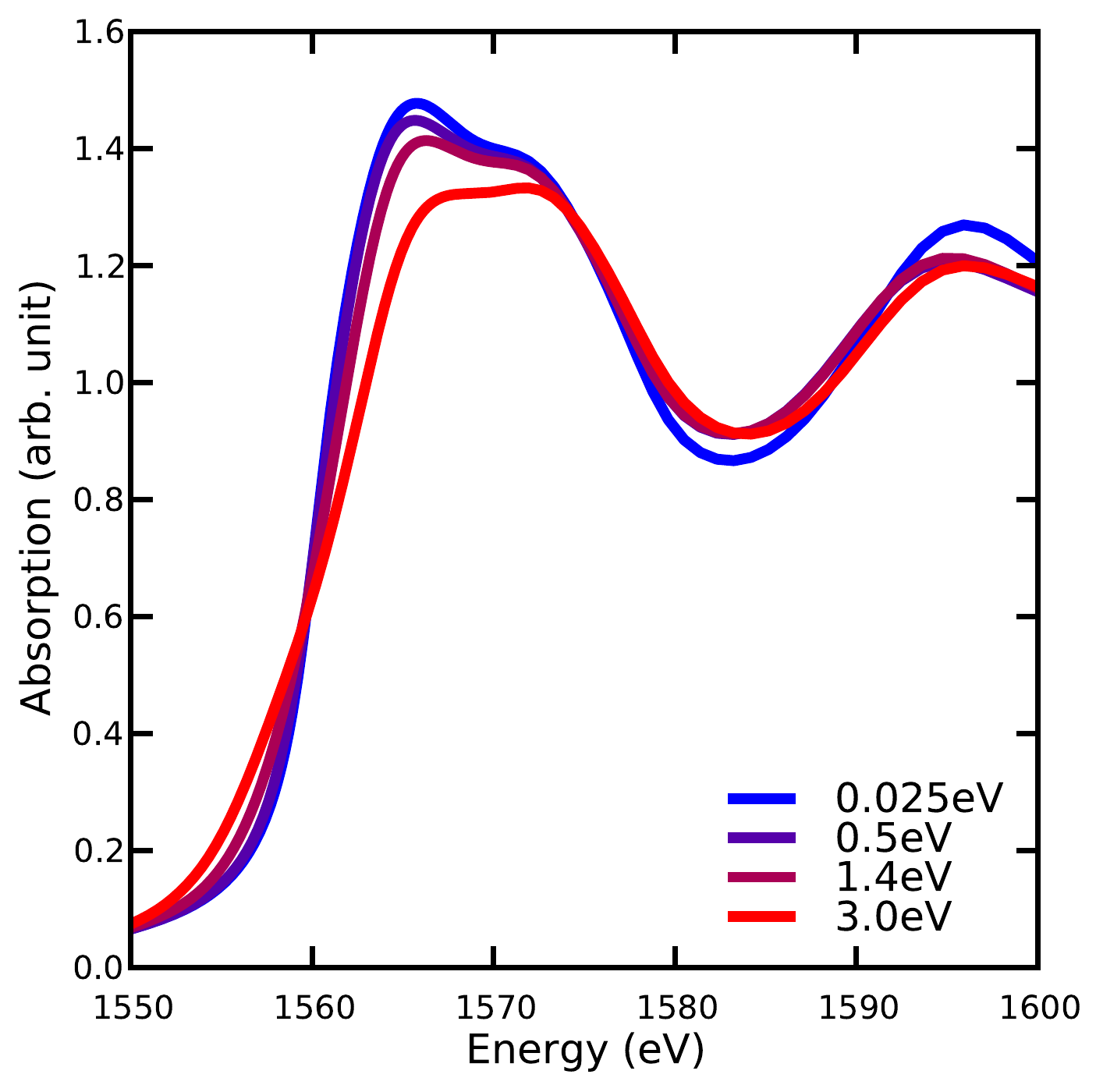}
        \caption{$\textnormal{K}$-edge XAS for aluminum vs electron temperature including only temperature dependent KSDT exchange correlation potentials (see text). Note that the edge broadens considerably with increasing temperature.}
        \label{fig:al_xas_electron}
    \end{figure}

     \begin{figure}[t]
        \centering
        \includegraphics[width=0.40\textwidth]{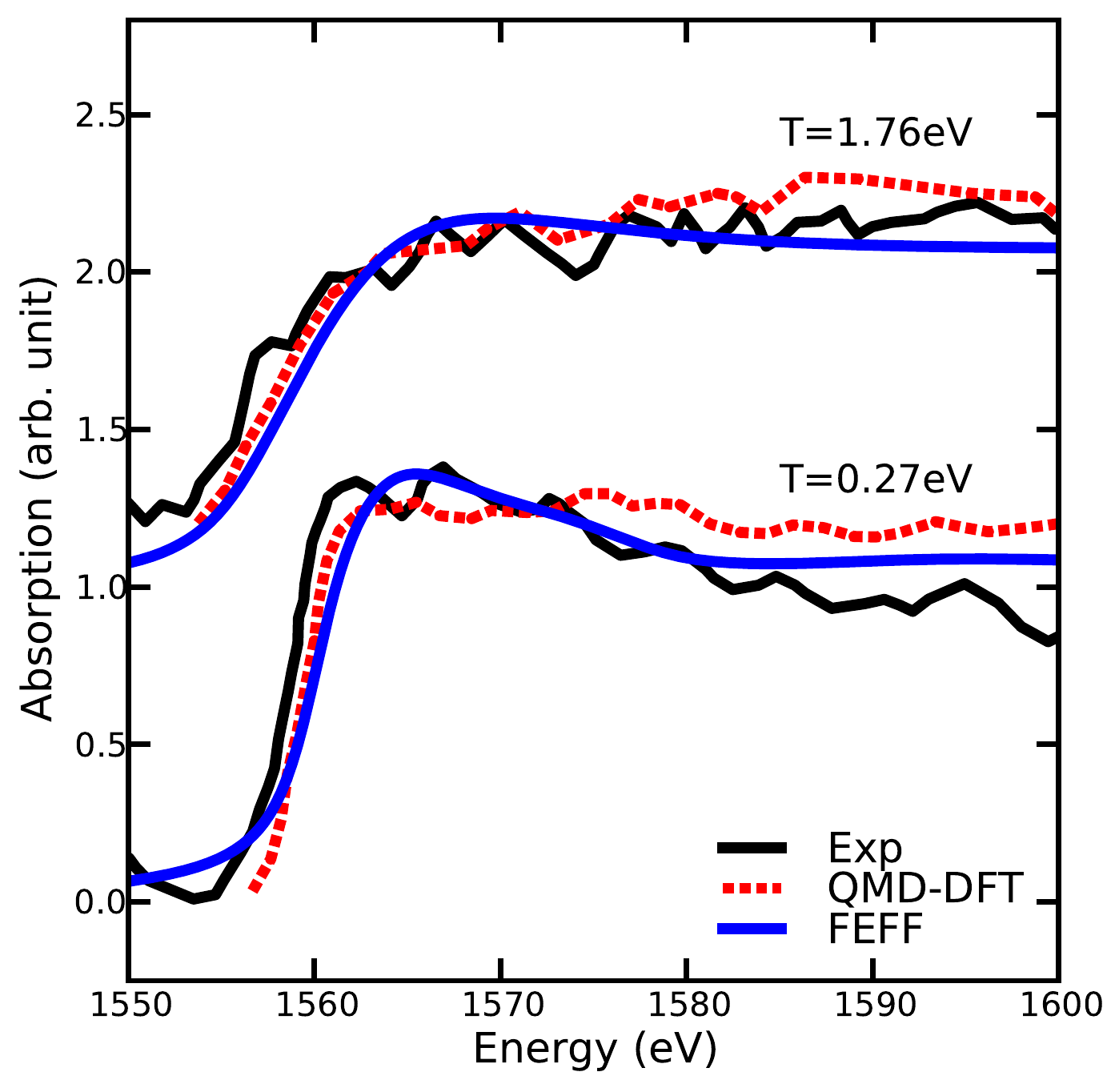}
        \caption{Calculated K-edge XAS  for equilibrium Al ($T_e =T_L$) at solid density vs
        %(Top) Experimental absorption spectra at $T=$ 0 and 0.09 eV.\cite{Mancic2010} FEFF simulation are done at the same temperature as the experiment. From bottom to top, QMD-DFT are simulated at $T=$ 0 and 0.07 eV while HNC-NPA are simulated at $T=$ 0 and 0.1 eV.  (Bottom)
        % {\bf \textcolor{red}{[TUN: PLEASE REMOVE CURVE at T=1.4 - it doesn't add much since you also show 1.76]}}
  experimental XAS at $T=$ 0.27 and 1.76 eV. From bottom to top, QMD-DFT are simulated at $T=$ 0.43, 0.86 and 0.07 eV while FEFF simulation are done at $T=$ 0.27, 1.4 and 3.0 eV. Note the substantial broadening of the edge at increased $T$, roughly consistent with experiment, and that lattice vibration effects completely damp the fine-structure. The fluctuations in the experimental XAS reflect experimental noise.}
        \label{fig:al_xas}
    \end{figure}
    Next, we investigate the temperature dependence of the K-edge XAS of Al at normal density up to WDM temperatures. We show the calculated K-edge XAS for electron temperatures $T_e$ = 0.025, 0.5, and 3.0 eV in Fig.\ \ref{fig:al_xas_electron}. The behavior at the edge is consistent with broadening of the Fermi-Dirac distribution due to the nearly free electron density of states of Al. The ``pre-edge" behavior is due to the increasing contribution from previously occupied states below the $T=0$ Fermi level.
% as $T$ is increased.

    In order to account for thermal vibrations, we use the correlated Debye model ($T_{D}$ = 430 K. For comparison we show DFT based XAS calculated using  quantum molecular dynamics (QMC) averaged over several QMD runs \cite{Mancic2010}.
    %For the HMNC, the neutral pseudo atom (HNC-NPA) model was used \cite{Peyrusse08, Peyrusse10}. This model assumes that the ions are surrounded by a spherical electron cloud forming a locally neutral sphere, and is effective for computing properties of hot dense plasma, where QMD is prohibitively expensive computationally.
    Investigations of the Al K-edge under isochoric heating  ($T_e =T_L$) at normal density were also done,\cite{Mancic2010} with measurements  at temperatures $T_e=T_L=$ 0.025, 0.09, 0.27, 1.40 and 1.76 eV, as  in Fig.\ \ref{fig:al_xas}.
    %This work showed that    the QMD-DFT approach agrees better with   experiment at low temperature ($T <$ 0.4 eV), while HNC-NPA is  better at higher temperatures.
    Our calculation with the correlated Debye model  agrees with fairly well with these results.
    %. At lower temperatures, our model also yields a good approximation to the more computationally demanding QMD-DFT method.
    In principle, the correlated Debye model breaks down at high temperatures ($T> 0.27$ eV) when anharmonicity  becomes large, but this is not a serious problem as the   fine-structure in the XAS is largely suppressed at these high temperatures.

\subsection{Warm Dense Cu}

% As a second example, we present results for the XANES spectra of warm dense copper. TODO: These results show that the computed spectra follows the energy shift of the experiment.

    Being a noble transition metal, Cu has substantially different excited-state properties compared to simple nearly free electron metals like aluminum, due  to the highly localized $d$-band just below the Fermi level. Hence, at elevated temperatures, the   XAS also differs. Several works \cite{Cho11, Cho2016, Mahieu2018, Jourdain18} studied these  changes in the $L_{2,3}$-edge of Cu XAS in WDM using a more computationally expensive AIMD simulation. Here we investigate the XAS of Cu using our SCF RSGF approach and the correlated Debye model.
     \begin{figure}[t]
        \centering
        \includegraphics[width=0.40\textwidth]{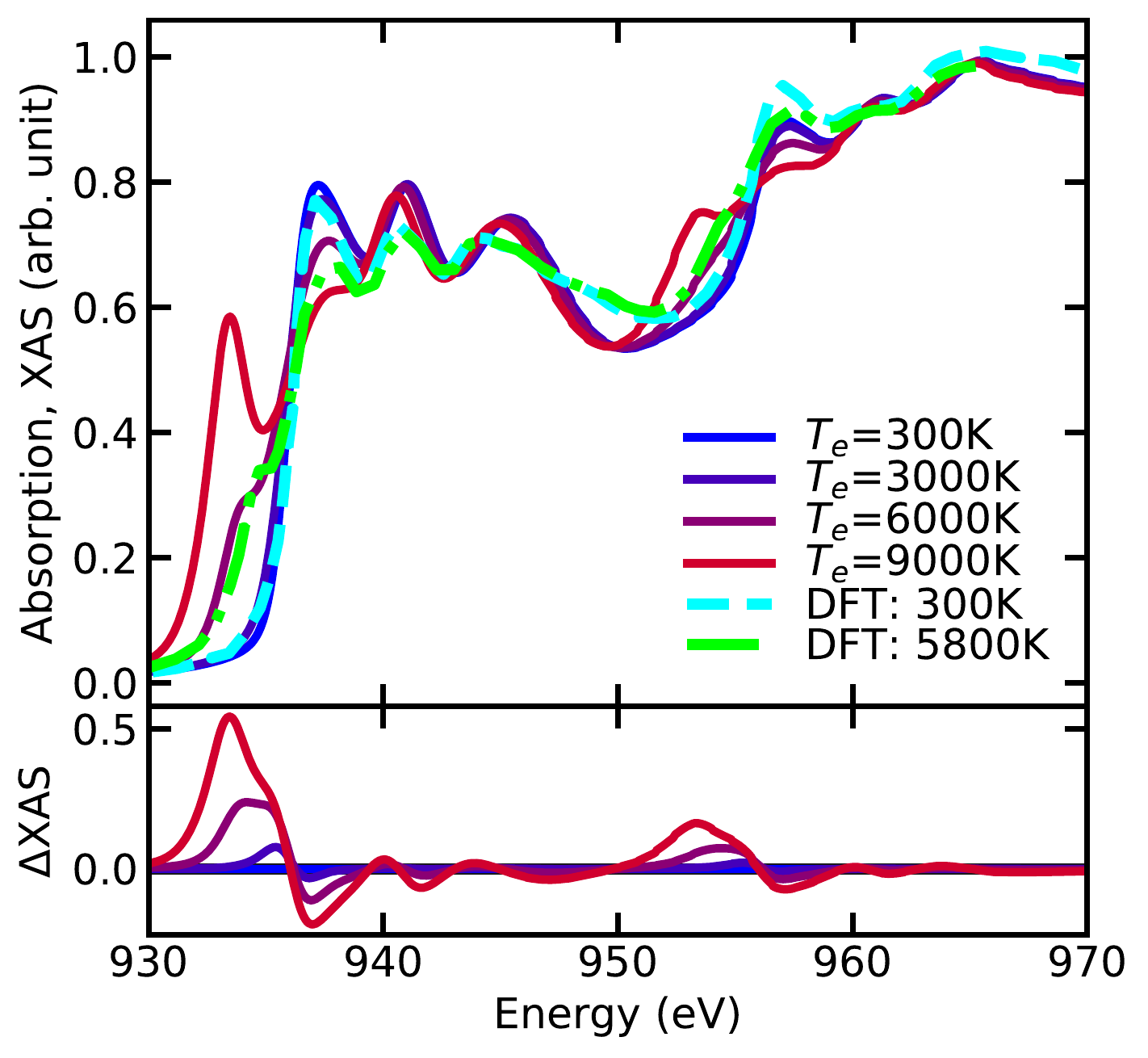}
        \includegraphics[width=0.40\textwidth]{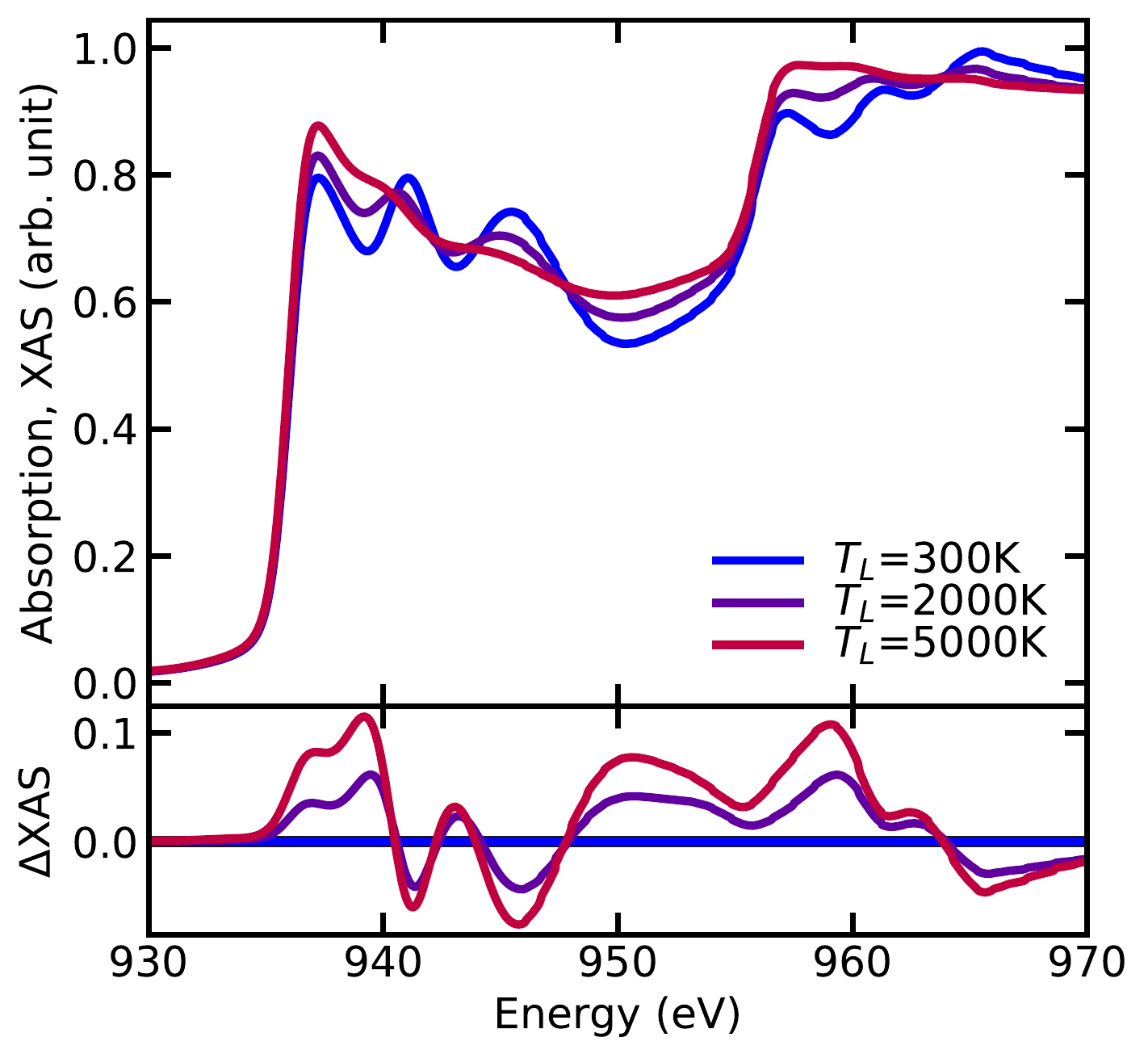}
        \caption{$\textnormal{L}_3$-edge XAS of Cu at normal density for (Top) electron temperature $T_e=$ 300 K up to 9000 K at lattice temperature $T_L=$ 300 K;  and (Bottom) for lattice temperature $T_L=$ 300 K up to 5000 K at electron temperature $T_e=$ 300 K. The DFT calculation
        %by \citeauthor{Jourdain20}
        \cite{Jourdain20} is shown for $T_e=$ 300 K (light blue) and 5800 K (green). The lower panels show the change in XAS with respect to  normal conditions $T_e = T_L = $ 300 K.}
        % TUN: PLEASE REMOVE DELTA XAS PLOT IN BOTTOM}
        \label{fig:cu_xas_temp}
    \end{figure}
    Fig. \ref{fig:cu_xas_temp} shows the simulated XANES spectra at normal density for various temperatures. Our calculation is broadened with a Gaussian to match that in the the DFT-based calculation \cite{Jourdain20}. As expected, the pre-edge structure in the XAS can be attributed to the increasing contribution from the d-states as the Fermi-Dirac distribution broadens with increasing $T$ \cite{Cho11, Cho2016,Mahieu2018, Jourdain18}. Conversely, states just above the Fermi level have a reduced contribution leading to a smaller peak just above the edge. The changes in XAS due to electronic temperature are mainly localized in the near edge,  whereas that due to the lattice temperature affects the region above the edge, especially the fine structure.
    \begin{figure}[t]
        \centering
        \includegraphics[width=0.4\textwidth]{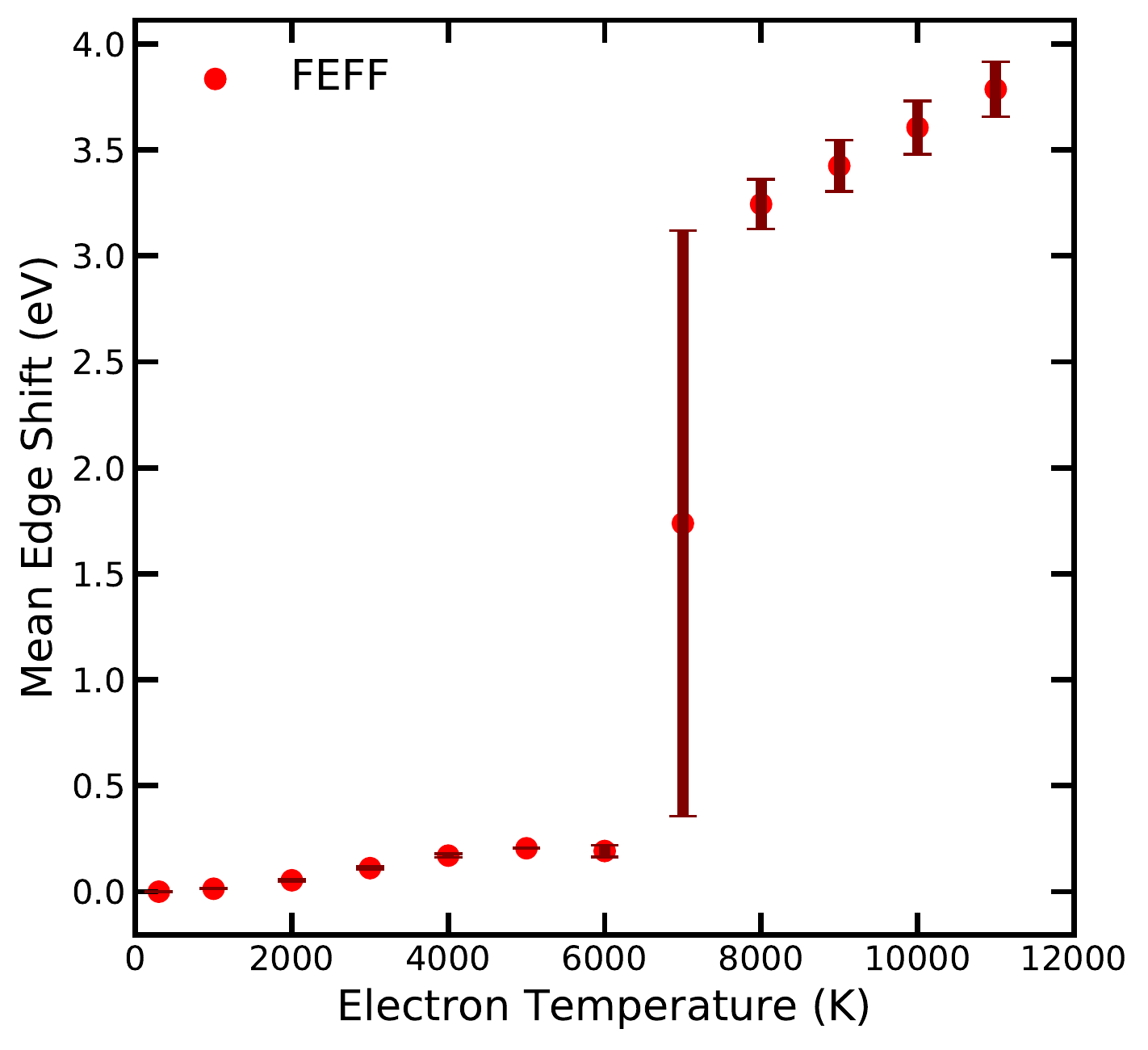}
        \includegraphics[width=0.4\textwidth]{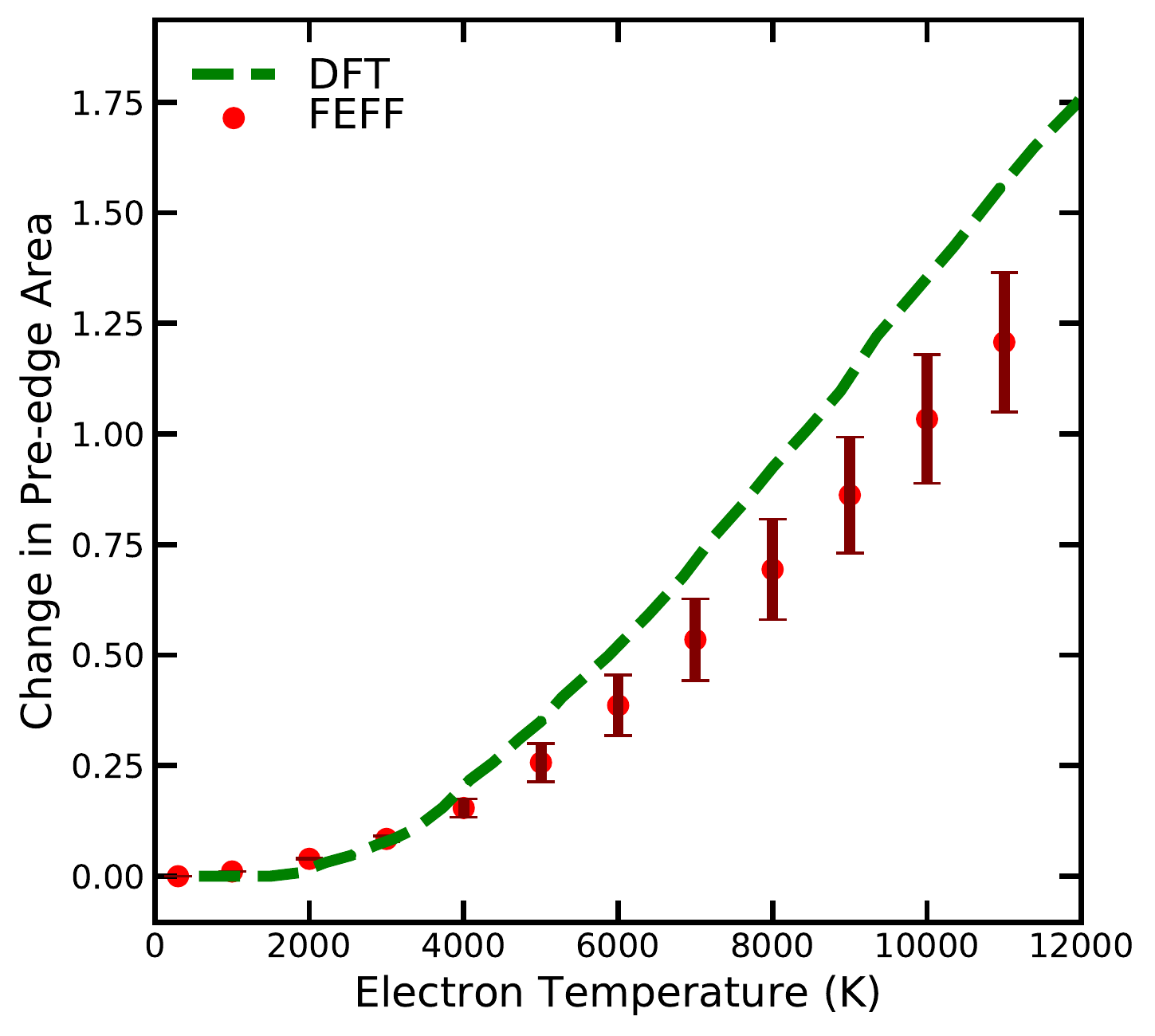}
        \caption{(Top) Edge shift in the $\textnormal{L}_3$-edge of Cu as a function of electron temperature (red); the sudden shift is due to the onset of the $d$-band edge.
        (Bottom) Change in pre-edge area as a function of electron temperature. The DFT calculation (green dashes)  \cite{Jourdain20} was computed using ABINIT at lattice temperature $T_L$ = 300 K. Note that in constrast to the edge shift, the pre-edge area varies smoothly with temperature for Cu.}
        \label{fig:cu_features}
    \end{figure}

     \begin{figure}[h!]
        \centering
        \includegraphics[width=0.4\textwidth]{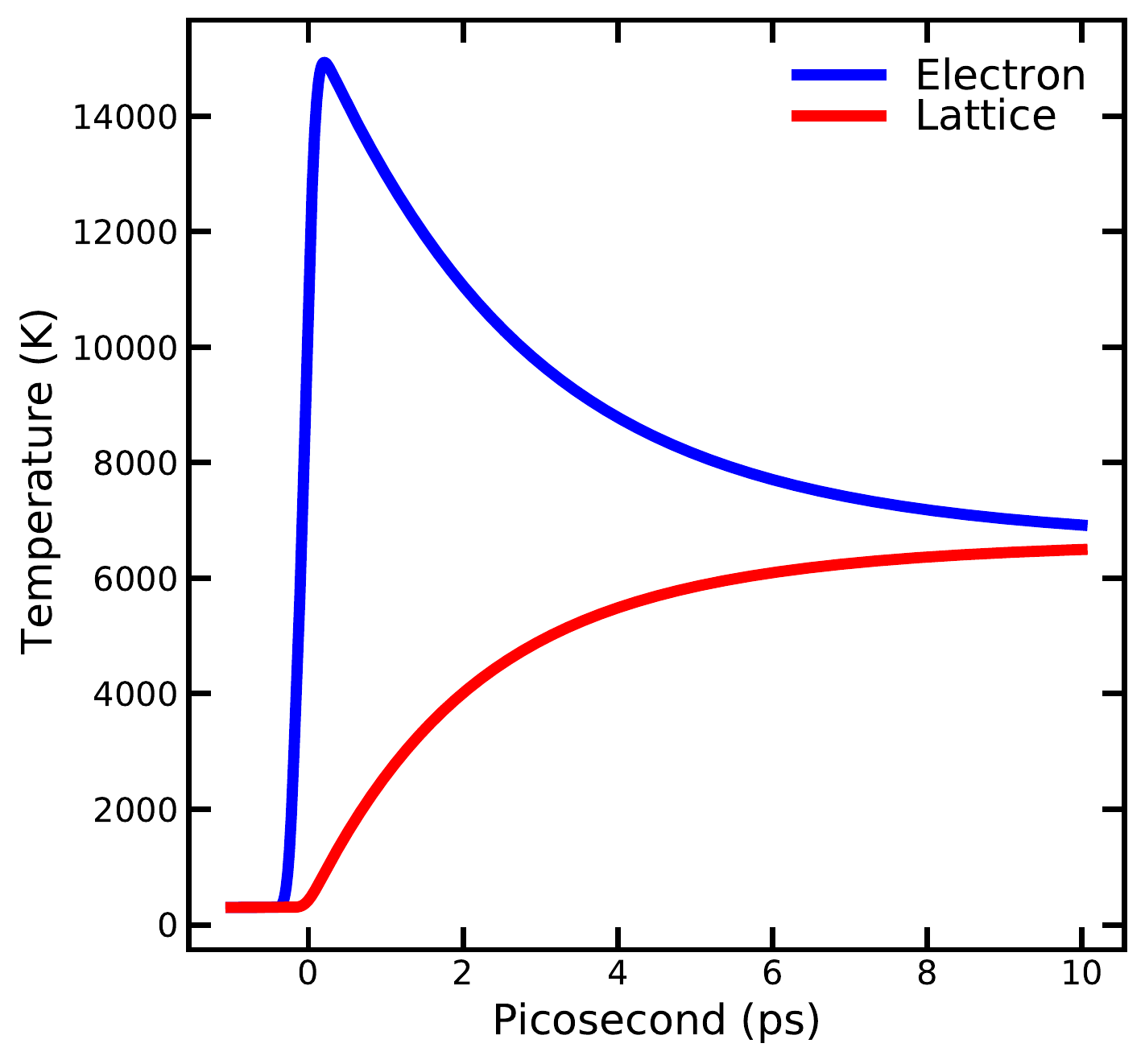}
        \caption{Dynamics of electron temperature (blue) and lattice temperature (red) based on two-temperature model calculation. The absorption power is calculated using Beer-Lambert law for a Gaussian profile 400 nm laser source for a 70 nm thick copper.}.
        \label{fig:cu_ttm}
    \end{figure}
    The edge-shift and the pre-edge area are both candidates for assessing the internal temperature of a system. In order to extract the edge shift and pre-edge area, the edge position is defined to be the first local maximum of the first derivative in the reference XANES spectrum. The electron temperature dependence of the edge shift and pre-edge area are shown in Fig.\ \ref{fig:cu_features}. Note that the pre-edge area is monotonically increasing with the electron temperature and it becomes linearly correlated to $T_e$ above 6000 K, consistent with previous results.\cite{Jourdain20}
    In contrast the edge-shift is highly non-linear, exhibiting an abrupt shift at about 7000 K. Therefore, for Cu and probably for other transition metals as well, the pre-edge area is a better proxy for an electron temperature thermometer, compared to the edge shift.

\begin{figure}[t]
        \centering
        \includegraphics[width=0.4\textwidth]{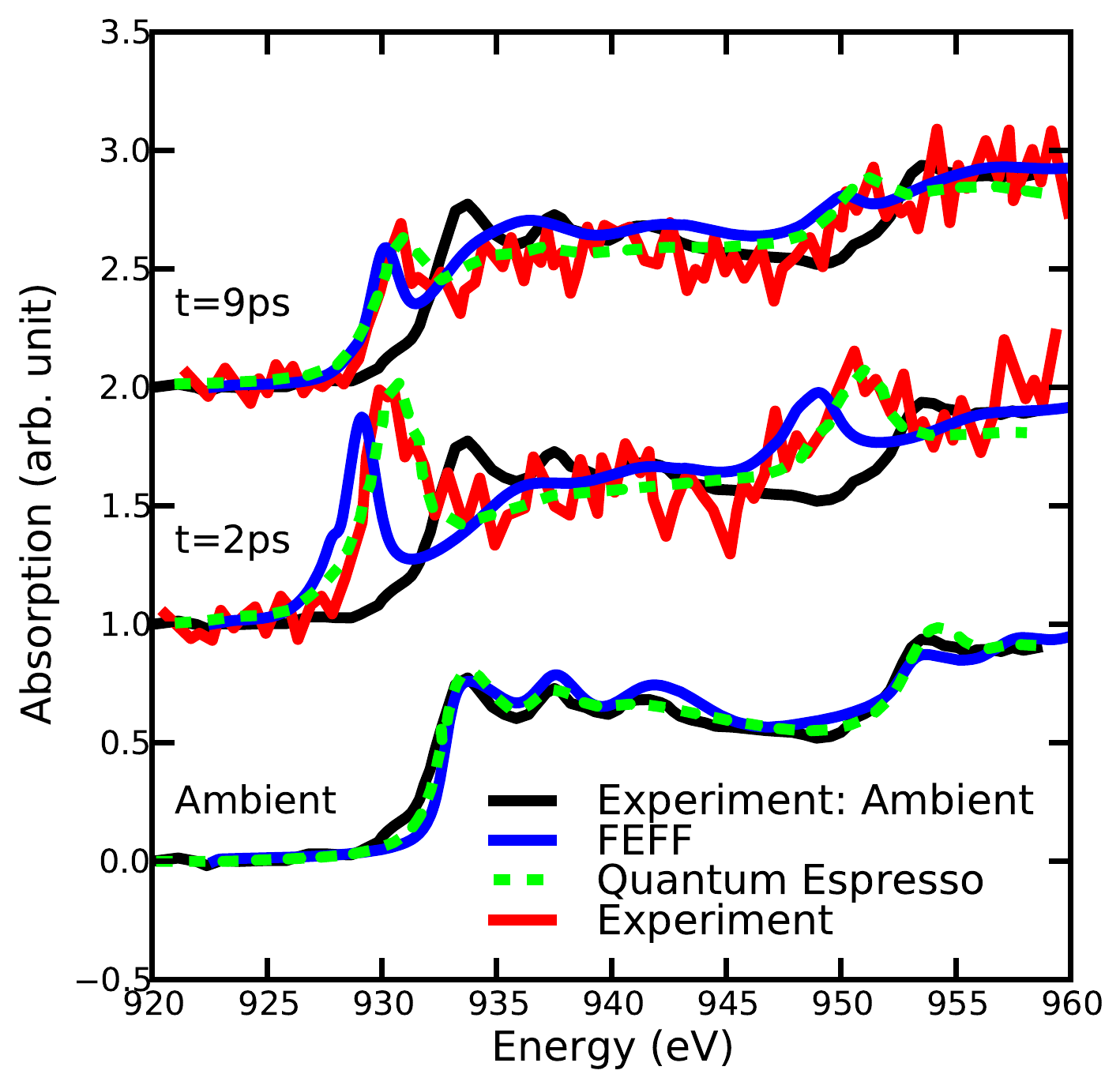}
        \caption{Time-resolved XANES for 70 nm copper at different delay-times $t=$2 ps and 9 ps. The experimental measurement (red) is averaged over 150 snapshots. The ambient condition measurement is averaged over 70 ps. Our calculations (blue) for different times are computed at $T_e=$ 10200 K and 6000 K. The DFT results\cite{Cho11} (green) are calculated using the Quantum Espresso code at the same temperatures.}
        \label{fig:cu_xas}
    \end{figure}

    Next, we compare our finite-temperature calculation to time-resolved, WDM experiments \cite{Cho11}. The system being explored is a 70 nm copper foil heated optically by a 400 nm laser at a fluence of 0.33 J/$\textnormal{cm}^2$. As a consequence the electrons are optically excited leading to an initial  huge pre-edge peak below the $L_3$-edge. In order to model the temperature evolution, we used a  two-temperature (2T)  model.\cite{Cho11} with the same parameters for the electron heat capacity and electron-phonon coupling factor as in Zhibin et al. \cite{Zhibin08}.
    The temperature evolution is shown in Fig.\ \ref{fig:cu_ttm}. Note that the lattice temperature raises quickly above the melting temperature $\sim$ 1358 K  under 1 ps due to the strong electron-phonon coupling in copper.
    %{\bf {[TUN: PLEASE ADD BACK 2-T model figure; also maybe include the 2-T model equations ?]}}
    Our simulations use XAS from 20 atomic configurations taken from QMD calculations using  the VASP  code \cite{Kresse96, Kresse99} with generalized gradient approximation (GGA) exchange-correlation potentials \cite{Perdew96}. We also used the PAW potentials with an energy-cutoff of 590 eV. The system with a 2$\times$2$\times$2 supercell constructed from a conventional unit cell of 8 atoms
    %\textcolor{red}{[HOW MANY ATOMS ARE IN THIS CELL? FCC PRIMITIVE CELL HAS ONLY 1 ATOM, SO 2x2x2 WOULD ONLY HAVE 8 ATOMS. A CONVENTIONAL CELL HAS 4 ATOMS, SO 32. STILL PRETTY SMALL.]}
     is propagated  with a time step of 1 fs to reach equilibration, and the sampling of configurations is performed by randomly sampling from a 2 ps long trajectory with a time step of 1 fs. We compare our simulation at  temperatures $T_e$ = 300 K, 10200 K and 6000 K for the time-delays $t<$ 0 ps, $t$= 2 ps and $t$= 9 ps respectively, with those from  DFT-based calculations\cite{Cho11} in Fig \ref{fig:cu_xas}. Note that our  calculation underestimates the Fermi level for $T_e$ = 10200 K ($t=2$ ps) by a few eV, leading to the observed shift in the XANES.
     %\textcolor{red}{[IS THIS UNDERESTIMATE OF THE FERMI-LEVEL DUE TO AN ERROR IN THE POSITION OF THE D-STATES?]}

     \begin{figure}[b]
        \centering
        \includegraphics[width=0.4\textwidth]{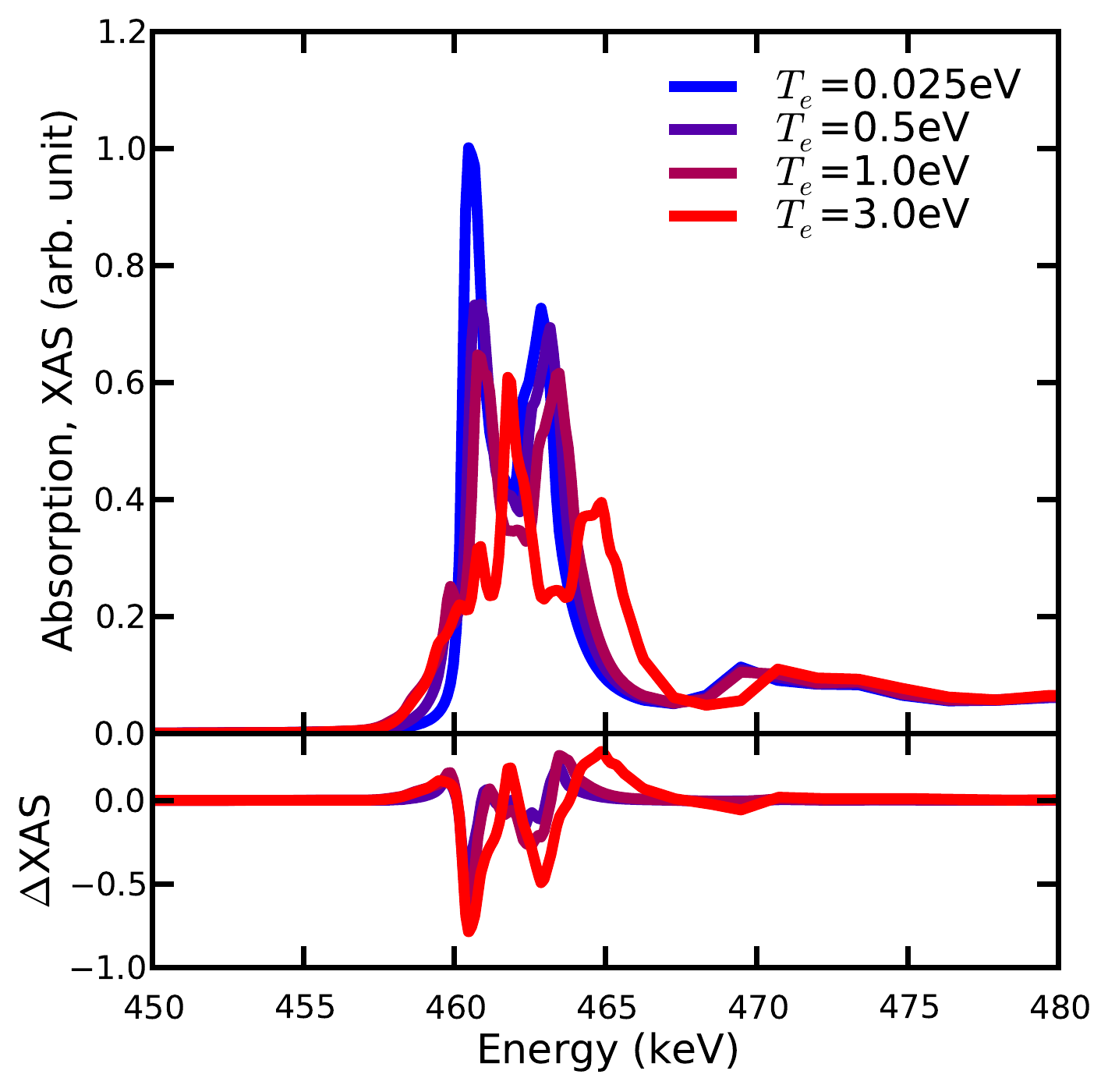}
        \includegraphics[width=0.4\textwidth]{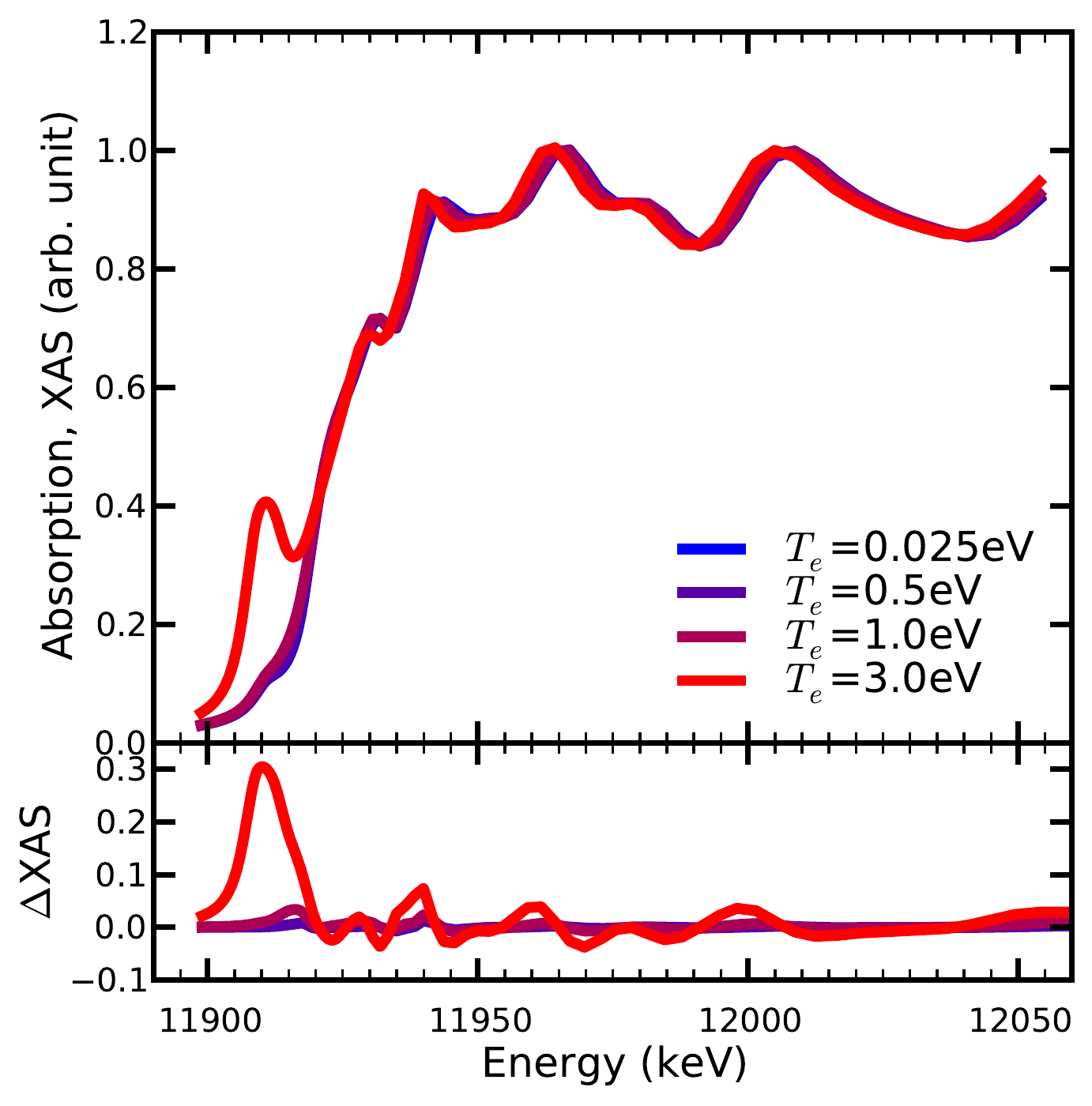}
        \caption{$\textnormal{L}_3$-edge absorption for titanium (Top) where the structure reflects that of the unfilled d-bands, and gold (Bottom) where the d-bands are nearly filled. The difference ${\Delta}$XAS between the XAS at finite $T$ and zero temperature is also shown.}
        \label{fig:ag_ti_xas}
    \end{figure}

\subsection{Transition Metals: Ti and  Au}
   As examples of other FT $L_3$-edge XAS calculations, we present FT calculations for a an early transition metal (titanium), where the $d$-bands are partially filled, and for a late transition metals (gold) where the $d$-bands are full. Fig.\ \ref{fig:ag_ti_xas}. shows the temperature dependence of the L-edges of these materials.
   The XANES of Ti is blue-shifted with increasing temperature because the chemical potential is located in the middle of the $d$-band and the density of states above the chemical potential is higher. Also, the broadening of the Fermi-Dirac distribution includes a wider energy range of $d$-states in the transition, which affects  the onset of the pre-edge. In contrast, the XANES of gold (Au) is red-shifted, as shown in bottom plot of Fig.\ \ref{fig:ag_ti_xas}. This opposing behavior is due to the higher density of states below the chemical potential. Again, the onset of the pre-edge structure is due to the broadening of the Fermi-Dirac distribution and shift of the chemical potential.

\subsection{MgO}

\begin{figure}[h]
        \centering
        \includegraphics[width=0.4\textwidth]{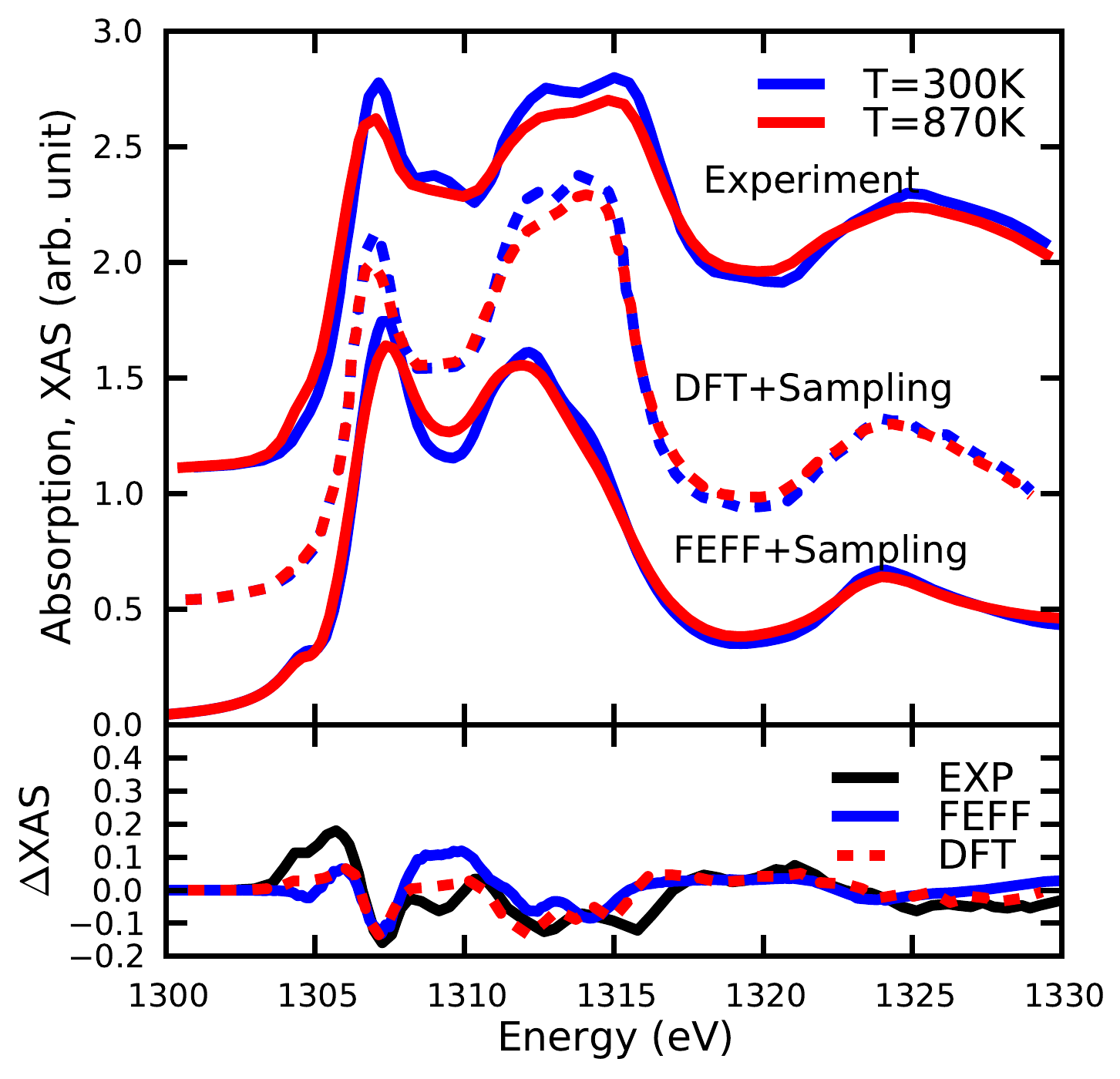}
        \caption{Mg K-edge spectra for MgO at temperature T=300 K (blue) and 870 K (red). The top plot shows the experimental spectra\cite{Nemausat15}, the DFT spectra \cite{Nemausat15} and the FEFF quasi-harmonic spectra. The bottom plot shows the spectra difference with respect to T=300K for the experiment (solid black), FEFF (solid blue) and DFT (dashed red) are shown.}
        \label{fig:mgo_example}
    \end{figure}

    As a final example, we present results for the Mg K-edge XANES of MgO with FT lattice behavior calculated using the quasi-harmonic approximation. Fig.\ \ref{fig:mgo_example} shows the average XAS for 30 randomly sampled configurations at equilibrium temperatures $T_e = T_l = $ 300K and 870K. The configurations are obtained by sampling the normal modes calculated from dynamical matrices at the experimental lattice constants and the given temperatures using Quantum Espresso code \cite{QE-2009, QE-2017, doi:10.1063/5.0005082}, with GGA functional\cite{qe_pseudo}  for a   16-atom  2 $\times$ 2 $\times$ 2 supercell constructred from a primitive cell. The electronic density of state is computed with a 6 $\times$ 6 $\times$ 6 k-point grid with an energy cutoff of 58 Ry. Next, we use the density-functional perturbation theory to compute the dynamical matrix calculation at the $\Gamma$-point. The long range electric fields is accounted for from the calculation of the Born effective charge.

     The FEFF calculations are found to overestimate the screening effect of light element oxides \cite{Nakanishi_2009}. Instead of using the ad-hoc $Z+1$ approach to overcome the strong screening effect of the final state rule, we used the random-phase approximation (RPA) for the core hole screening in FEFF. The RPA is more accurate in this case, being similar to a Bethe-Salpeter equation. \cite{Ankudinov03, Rehr05} Finally, we add an additional energy shift of 2.6 eV to the chemical potential. As a comparison, we show the experiment and DFT calculation by \citeauthor{Nemausat15} using the stochastic self-consistent harmonic approximation\cite{Nemausat15}  in Fig.\ \ref{fig:mgo_example}. Our muffin-tin potential model predicts a smaller excitations between 1310 eV and 1315 eV.

\section{CONCLUSIONS}

We have extended the real-space Green's function theory of XAS \cite{RehrAlbers2000} to finite temperatures (FT) up to the warm-dense matter regime $T\approx T_F$.  Our FT generalization  takes into account both electron temperature and lattice temperature effects.
 Briefly, a FT SCF procedure is carried out in the complex energy plane in terms of the FT one-electron Green's function. This builds in the FT exchange-correlation potential, using the KSDT parameterization. The FT self-energy is also important for XAS, since it accounts for final-state effects in the spectra including  corrections to shifts and broadening.  While dynamic exchange effects can also be included, due to the difficulty calculating the FT self-energy, a more efficient FT COHSEX approximation has been used.
 An important  difference from the $T=0$ theory of the XAS is an account of  smearing of the Fermi level due to FT occupation numbers. Vibrational effects are included in terms of a correlated Debye model for the fine-structure at low $T$ in XAS or a configurational average for XANES at high $T$. The FT generalization  introduced here is implemented as an extension of the FEFF codes in a new version FEFF10 \cite{feff10ref}. 
 %\cite{Rehr09}, now permitting calculations of FT XAS   %applicable to systems throughout the periodic table.
 The approach has been  tested against various experiments, and typically yields good quantitative agreement. We believe these developments may be useful in interpretation of many experiments, e.g., for studies non-equilibrium behavior, extreme conditions, and shocked conditions.  The approach can also be used to differentiate between lattice and electron temperature effects. 
 \linebreak 
 
 \begin{acknowledgments}
 We thank R. Albers, B. Fultz, D. Prendergast, G. Seidler, S. Trickey, F. Vila, and H. Wende for comments and suggestions.  This work is supported by DOE BES Grant DE-FG02-97ER45623. The development of FEFF10 carried out within the Theory Institute for Materials and Energy Spectroscopies (TIMES) at SLAC, and  is supported by
 the U.S. DOE, Office of Basic Energy Sciences, Division of Materials Sciences
 and Engineering, under contract DE-AC02-76SF00515.
 \end{acknowledgments}

 \bibliographystyle{apsrev4-2}
%\bibliography{reference}

\end{document}